\documentclass[conference]{IEEEtran}
\IEEEoverridecommandlockouts
\usepackage{cite}

\usepackage{xcolor}
\usepackage[colorlinks=true, linkcolor=red, urlcolor=red]{hyperref}
\usepackage{orcidlink}

\usepackage{amsmath,amssymb,amsfonts}
\usepackage{algorithmic}
\usepackage{graphicx}
\usepackage{textcomp}
\usepackage{xcolor}
\def\BibTeX{{\rm B\kern-.05em{\sc i\kern-.025em b}\kern-.08em
    T\kern-.1667em\lower.7ex\hbox{E}\kern-.125emX}}

\begin{document}

\title{Adaptive Real-Time Scheduling Algorithms
for Embedded Systems}

\author{\IEEEauthorblockN{1\textsuperscript{st} Abdelmadjid Benmachiche\orcidlink{0000-0002-0690-2625}}
\IEEEauthorblockA{\textit{Department of Computer Science} \\
\textit{LIMA Laboratory}\\
\textit{Chadli Bendjedid University}\\
El-Tarf, PB 73, 36000, Algeria\\
benmachiche-abdelmadjid@univ-eltarf.dz}

\and
\IEEEauthorblockN{2\textsuperscript{nd} Khadija Rais\orcidlink{0009-0004-3907-7782}}
\IEEEauthorblockA{\textit{Informatics and Systems (LAMIS)} \\
\textit{Echahid Cheikh Larbi Tebessi University}\\
Tebessa, 12002, Algeria \\
khadija.rais@univ-tebessa.dz}

\and
\IEEEauthorblockN{3\textsuperscript{rd} Hamda Slimi\orcidlink{0000-0002-8494-6551}}
\IEEEauthorblockA{\textit{Informatics and Systems (LAMIS)} \\
\textit{Echahid Cheikh Larbi Tebessi University}\\
Tebessa, 12002, Algeria \\
slimi.hamda@univ-tebessa.dz}
}

\maketitle

 \IEEEpubid{%
  \begin{minipage}{\textwidth}
    \vspace{7em}
    \makebox[0pt][l]{\textbf{The second National Conference on Artificial Intelligence and Information Technologies}-\textit{ Chadli Bendjedid El-Tarf University - EL TARF (Algérie)}
     }
  \end{minipage}%
}

\begin{abstract}
Embedded systems are becoming more in demand to work in dynamic and uncertain environments, and being confined to the strong requirements of real-time. Conventional static scheduling models usually cannot cope with runtime modification in workload, resource availability, or system updates. This brief survey covers the area of feedback-based control (e.g., Feedback Control Scheduling) and interdependence between tasks (e.g., Symbiotic Scheduling of Periodic Tasks) models. It also borders on predictive methods and power management, combining methods based on Dynamic Voltage and Frequency Scaling (DVFS). In this paper, key mechanisms are briefly summarized, influencing trade-offs relating to adaptivity/predictability, typical metrics of evaluation, and ongoing problems, especially in situations where safety is a critical factor, giving a succinct and easy-to-understand introduction to researchers and practitioners who have to cope with the changing environment of adaptive real-time systems.
\end{abstract}
\begin{IEEEkeywords}
Adaptive scheduling, real-time embedded systems, feedback control scheduling, symbiotic task scheduling, DVFS
\end{IEEEkeywords}

\section{Introduction}

The embedded systems that are used in modern times, from automated vehicles to smart medical equipment, are increasingly dynamic environments where the workloads, availability of resources, and external conditions vary dynamically. Conventional real-time scheduling algorithms, such as Rate-Monotonic (RM) and Earliest Deadline First (EDF), make an additional assumption that the set of tasks is known and that the execution times are fixed, where they cannot cope with uncertainty at runtime without making timing guarantees \cite{datta2022adaptive}.

\textbf{In autonomous systems, such as next-generation robots, adaptive path planning algorithms like hybrid PSO-APF approaches \cite{benmachiche2025adaptive} demonstrate the need for real-time adaptation in dynamic environments with static and dynamic obstacles. Similarly, hybrid BFO/PSO algorithms for mobile robot navigation \cite{makhlouf2024enhanced} showcase how bio-inspired optimization can enhance real-time decision-making in embedded systems.}

To bridge this gap, adaptive real-time scheduling has gained significant traction, enabling systems to modify priorities, execution budgets, or resource allocations at runtime based on feedback, prediction, or environmental sensing. Among the promising directions are feedback control scheduling (FCS) frameworks that use control-theoretic loops to maintain performance under load variations \cite{lu2002feedback}, and machine learning–enhanced predictive schedulers that leverage historical data to anticipate task behavior and proactively adjust schedules \cite{Gracias}. The integration of adaptation based on machine learning into on-board safety-critical systems is of particular importance. Furthermore, the detection of real-time anomalies of onboard aircraft devices requires that the timing and accuracy of detection be optimized for timing and accuracy while respecting the strictest limits of energy consumption and latency constraints \cite{ben}.

Recent research highlights this development, for instance Subramaniyan et al. proposed FC-GPU \cite{subramaniyan2025fc}, a real-time embedded feedback control scheme that dynamically controls the allocation of GPU resources to soft deadline-based real-time edge AI applications, an essential feature of edge AI applications. In the meantime, Goksoy suggested a runtime monitoring system of ML-based schedulers that indicates a change of distribution during the execution of the task and allows a safe adjustment of safety-critical environments \cite{goksoy2024runtime}. In another study \cite{zhao2024task}, Zhao revealed a task-degradation-aware adaptive scheduler that is effective in managing the dynamically applications and maintaining the real-time restrictions.

This survey provides an introduction to adaptive real-time scheduling strategies of embedded systems, especially the feedback-based, predictive, and hybrid methods. We discuss how they can be integrated with system mechanisms such as DVFS, some of the most significant trade-offs (e.g., adaptivity vs. predictability), and the current gaps in verification and benchmarking.

\section{Adaptive Scheduling Approaches}
\subsection{Feedback-Based Scheduling}
Adaptive scheduling reacts to real-time system performance metrics (e.g., deadline miss ratio, queue length, execution delays, or resource utilization) based on feedback to dynamically modify CPU/GPU allocation, task priorities, or execution budgets. Control-theoretic principles are used to determine the feedback-based reaction to the tactical choices made by the adaptive scheduling algorithm. In contrast to traditional methods of open-loop scheduling, which are based on offline worst-case execution time (WCET) measurements, feedback-based scheduling performs continual monitoring of system behavior and a corrective response in the form of control actions to ensure system stability, timing constraints, and quality of Service (QoS), and it does not require accurate a priori knowledge of workload.

One of the foundational structures in this area was presented by Lu et al. \cite{lu2002feedback} with the concept of Feedback Control Scheduling (FCS), the application of classical control theory to real-time systems. Their work enabled the systematic modeling of computing systems as dynamical systems and the development of controllers to control ratios of CPU utilization and deadline misses. This paper showed that analytically tuned feedback controllers can offer high levels of transient and steady-state performance even in cases where task execution times change by large factors at runtime.

In another research paper by  Subramaniyan and Wang \cite{subramaniyan2025fc}, FC-GPU (Feedback Control GPU Scheduling) is introduced, the first feedback-based scheduling model specifically developed to fit GPUs in embedded and real-time systems. They use a multi-input, multi-output (MIMO) system to model the contention of resources in GPUs to extend feedback control scheduling to the heterogeneous architecture of a GPU. A MIMO controller is a dynamically adjusted task invocation rate controller, which adjusts the rate in response to measured response times. According to the experimental findings on NVIDIA RTX 3090 and AMD MI-100 GPUs, both show considerable enhancements in real-time and high-endurance performance under fluctuations in workloads during runtime.

Pan and Wei proposed a real-time workflow feedback scheduling system in cloud and distributed systems using deep reinforcement learning. They apply a regularized Deep Q-Network (R-DQN) that assigns workflow tasks to virtual machines depending on the condition of the system. This architecture proves to be more adaptive, scalable, and robust than conventional heuristic and fixed schedulers, which essentially serve as a smart feedback controller that learns the best scheduling policies online \cite{pan2024deep}.

Scheduling of industrial processes has also been done successfully using feedback principles. He et al. \cite{he2025closed} suggested a closed-loop gasoline blending scheduling scheme that unites real-time optimization and slack-based feedback. Their approach adds slack variables to the constraints of the processes and feeds actual deviations back into the scheduling model and re-optimizes production plans while accounting for actual deviations. Such a strategy enhances the similarity between theoretical scheduling and the real behavior of operations in a highly dynamic environment.

\subsection{Symbiotic / Task-Coupling Scheduling}

Task-coupling or symbiotic scheduling is a higher-level approach to scheduling that takes advantage of inter-task dependencies and correlation/co-activation patterns in order to enhance system efficiency and predictability. This method, instead of handling tasks as individual entities, considers them cooperative or symbiotic. Cooperation and execution of tasks may allow for resolving resource contention while also improving cache and memory locality and end-to-end timing guarantees. The model works especially well in cyber-physical systems, multi-sensor hard workloads, and highly parallel architectures, in which the interactions among tasks significantly affect system-wide performance.

Posluns and Jeffrey \cite{posluns2025symbiotic}, in their article about Symbiotic Task Scheduling and Data Prefetching, introduce a modern architectural view of the concept of symbiotic scheduling. They present the Task-Seeded Prefetcher (TSP) and Memory Response Task Scheduler (MRS), a joint hardware-software system, which allows task scheduling and memory subsystem behavior to cooperate. TSP acquires task-specific data access patterns and prefetches memory at the granularity of short-lived tasks, whereas MRS takes short-lived tasks as the source of prefetch status decisions. This symbiotic association achieves significant reductions in exposed DRAM latency and substantial performance gains in large-scale manycore processors.

Symbiotic scheduling was first investigated in the context of Simultaneous Multithreading (SMT) architectures in the form of the Symbiotic Job Scheduling (SOS) scheme by Snavely and Tullsen \cite{articlsnavelye}. SOS is a dynamic approach that finds workloads that run effectively as job co-scheduling combinations by sampling different combinations that may or may not work effectively. Through intelligent pairing of threads by using online sampling and a symbiosis-based scheduling phase, SOS proved to achieve better system throughput and lower response time by exploiting complementary usage of resources by threads.

Symbiotic concepts have been applied to real-time IoT workloads in distributed and edge-cloud environments. A semi-dynamic and real-time multiplexing algorithm of task scheduling in cloud-fog IoT systems was proposed by Abohamama et al. \cite{abohamama2022real}, where implicit task coupling is observed by optimization by means of permutation based on a modified genetic algorithm. The algorithm clusters and ranks tasks to maximize execution locality and resource fit between the fog and the cloud nodes, resulting in significant gains in makespan, latency, and failure rate for delay-sensitive IoT applications.

Recently, Kwon et al.       \cite{kwon2025real} applied the concept of task-coupling to Industrial IoT (IIoT)-based flexible manufacturing systems that involve human-machine interaction. Their model considers operator-induced variation of tasks as explicit coupling events and pre-calculates executable joint resource strategies that combine CPU frequency scaling, memory allocation, and edge/cloud offloading choices. At runtime, the scheduler alternates between these coupled plans on the basis of perceived interactions, providing strict adherence to deadlines and optimizing energy use and system responsiveness.

\subsection{Predictive / ML-Based Scheduling}

Predictive scheduling is a data-driven approach to scheduling that takes advantage of historical traces, machine telemetry, and real-time machine data to predict task performance, workload, and resource usage. It is in contrast to reactive scheduling mechanisms. Such methods are proactive by planning how systems will behave and can take measures that include dynamically scaling resources, reprioritizing tasks, or preemptive migration before situations arise in which a deadline is missed or performance is compromised. It is a powerful paradigm in very dynamic environments, such as cloud data centers, edge computing platforms, the Internet of Things (IoT), and autonomous cyber-physical systems.

Pan and Wei \cite{pan2024deep} designed a representative cloud-based model and suggested a deep learning–based reinforcement learning scheduling system in real-time workflow applications. Their architecture uses a regularized Deep Q-Network (R-DQN), which dynamically plans workflow tasks to virtual machine instances depending on the perceived system state. The approach performs well in optimizing workflows for workload uncertainty and increased resource usage in a heterogeneous workflow structure because it learns the best scheduling policies online, thus making it highly scalable.

In another study, Kesavan and his colleagues \cite{kesavan2025reinforcement} proposed a model named Secure Edge-Enabled Multi-Task Scheduling (SEE-MTS) in edge- and IoE-based settings that combines reinforcement learning and security- and energy-conscious scheduling. Their system integrates edge computing and encrypted task management, dynamic key generation and verification mechanisms, and reinforcement learning to reduce task completion time and energy utilization. The framework is energy-efficient and offers strong security and assurance, which demonstrates the appropriateness of ML-based scheduling for safety-critical and resource-constrained distributed systems.

Conceptually and in terms of systems-level, Guo \cite{guo2025machine} explored the combination of neural networks and machine learning methods with real-time scheduling. This paper points out how neurodynamic systems and reinforcement learning can be used to solve constrained optimization problems subject to time constraints in real time, and the architectural and safety issues of determinism, hardware accelerators, and certification in safety-critical workloads. The research points out the significance of co-designing learning algorithms along with real-time system guarantees. 

Related AI approaches in other domains include intrusion detection systems using deep learning \cite{sedraoui2024intrusion} and voice recognition platforms using convolutional neural networks \cite{benmachiche2022development}, which demonstrate the potential of ML techniques for real-time embedded applications. Furthermore, optimization techniques like bacterial foraging optimization (BFO) applied to Hidden Markov Models for speech recognition \cite{benmachiche2020optimization, benmachiche2019optimization} illustrate how adaptive algorithms can optimize system parameters in real-time applications.

Similarly, Gracias and Brooklyn \cite{Gracias} discussed AI and ML applications in predictive scheduling within real-time operating systems (RTOS). Their work uses supervised learning models to estimate the execution time of tasks, reinforcement learning to produce adaptive scheduling policies, and context-aware prioritization of tasks using neural networks. Their findings depict better latency, predictability, and throughput than conventional heuristic schedulers, particularly when dealing with extremely variable and unpredictable workloads.

\subsection{ Hybrid / DVFS-Integrated Scheduling}
Hybrid scheduling approaches combine task management with Dynamic Voltage and Frequency Scaling (DVFS) for the simultaneous optimization of power consumption, performance, and thermal limits in real-time embedded systems. These methods allow fine trade-offs between power consumption and deadline guarantees and jointly control the operating frequency and voltage of processing units, thus making them very suitable for battery-operated, resource-constrained, or thermally sensitive applications.

The researchers in \cite{li2024fidrl} implemented the Flexible Invocation-Based Deep Reinforcement Learning (FiDRL) framework for DVFS scheduling in embedded systems. FiDRL enhances regular DRL methods by embedding the invocation timing of the agent within its action space, which allows the agent to self-invoke judiciously to minimize the overall energy of the system, accounting for the energy overhead of the DRL agent. FiDRL allows inter- and intra-task DVFS scheduling and also uses a hybrid on/off-chip algorithm to train and deploy on an embedded platform, achieving a 55.1\% decrease in the energy overhead of agent invocation.

In another study \cite{li2025real}, the authors proposed a real-time data-driven hybrid synchronization framework to address heterogeneous demand–capacity synchronization (HDCS) in flexible manufacturing systems. Their approach integrates hierarchical real-time data feedback with a ticket-enabled queuing mechanism (GiMS) to ensure seamless coordination between planning, scheduling, and execution. The framework combines global optimization models with local adaptive control mechanisms, enabling rapid reactions to disturbances while preserving overall system optimality.

For flexible and multiskilled manufacturing systems with unpredictable demand, Xinyi Li et al. \cite{li2025energy} defined a real-time data-driven hybrid framework integrating planning, scheduling, and execution (PSE). A stratified real-time feedback data system with hybrid optimization is included in the framework to achieve efficiency and responsiveness of local and global performance simultaneously through the use of human and machine cooperation. Experimental results show cost-efficiency, timeliness, and good resource management, underscoring how hybrid scheduling with real-time data feedback optimizes energy, time, and operational performance in Industry 5.0.

\section{Discussion}
Adaptive real-time scheduling has evolved from a niche concept into a necessity for modern embedded systems operating in dynamic, uncertain, and resource-constrained environments. The surveyed approaches, feedback-based, symbiotic, predictive, and hybrid DVFS-integrated, each address different facets of runtime adaptability. Feedback control offers robustness through continuous system monitoring and corrective actions, making it suitable for environments with bounded but unpredictable perturbations. Symbiotic scheduling exploits structural relationships among tasks, improving predictability in cyber-physical and multi-sensor systems where tasks are inherently interdependent. Predictive and ML-based methods, particularly those leveraging deep reinforcement learning, promise high adaptability in complex, non-stationary workloads but often at the cost of determinism and explainability. Meanwhile, hybrid DVFS-integrated strategies demonstrate that energy and timeliness can be co-optimized, especially in edge and battery-powered devices. However, all these paradigms share a common tension: the trade-off between adaptivity and predictability. While adaptivity enhances responsiveness to change, it inherently weakens the strong timing guarantees that traditional real-time systems rely on for safety certification. This tension becomes critical in domains like automotive or medical systems, where even minor violations of timing constraints can have severe consequences. Thus, the core challenge lies not in enabling adaptation, but in doing so safely and verifiably.

\section{Research Gaps}
Despite significant progress, several critical gaps remain unresolved in the field of adaptive real-time scheduling. First, there is a notable lack of formal verification frameworks for adaptive schedulers. Most ML-based or feedback-driven approaches operate as black boxes, making it difficult to provide mathematical guarantees about schedulability under all possible runtime conditions. Second, current evaluation methodologies are fragmented and non-standardized; researchers use custom workloads, metrics, and platforms, which hinders fair comparison and reproducibility. Third, the overhead of adaptation mechanisms, whether computational, memory, or energy, is often underreported or assumed negligible, yet it can dominate system behavior in ultra-constrained embedded devices. Fourth, there is insufficient work on cross-layer co-design, where adaptation decisions jointly consider scheduling, memory management, thermal constraints, and communication (e.g., in networked embedded systems). Finally, most adaptive schedulers assume soft or firm real-time constraints; extending these techniques to hard real-time contexts, where deadlines must never be missed, remains largely unexplored, especially when adaptation itself introduces non-determinism.

\section{Future Directions}
Future research should focus on bridging the gap between intelligent adaptivity and rigorous real-time guarantees. One promising direction is the development of certifiable adaptive schedulers, where learning-based components are augmented with runtime monitors, fallback policies, or formal wrappers that enforce safety boundaries. Another avenue is the creation of open benchmarking suites for adaptive real-time systems, including standardized task sets, fault injection models, and metrics that capture both timing fidelity and adaptation efficiency. Additionally, integrating lightweight explainable AI (XAI) into predictive schedulers could improve trust and debuggability, enabling developers to understand why a scheduling decision was made and whether it aligns with system-level objectives. Furthermore, as embedded systems become increasingly heterogeneous (e.g., CPU-GPU-FPGA SoCs), future schedulers must support cross-architecture adaptation, dynamically partitioning workloads across processing elements while respecting end-to-end deadlines. Lastly, the emergence of human-in-the-loop embedded systems, such as collaborative robots or assistive medical devices, demands schedulers that can adapt not only to computational load but also to human behavior, variability, and safety preferences, opening a new frontier in context-aware real-time adaptation.

\section{Conclusion}
Adaptive real-time scheduling has become indispensable for modern embedded systems that must operate reliably in dynamic, uncertain, and resource-constrained environments. This survey has outlined four dominant paradigms: feedback-based control, which offers stability through runtime monitoring; symbiotic/task-coupling scheduling, which exploits inter-task relationships for improved predictability; predictive and machine learning–based methods, which enable proactive adaptation in complex workloads; and hybrid DVFS-integrated techniques, which jointly optimize energy efficiency and timeliness. While each approach brings unique strengths, they all grapple with the fundamental tension between adaptivity and predictability, especially critical in safety-critical domains where timing guarantees cannot be compromised.

Despite significant advances, key challenges remain, including the lack of standardized benchmarks, limited formal verification for learning-driven schedulers, and insufficient support for hard real-time guarantees under adaptation. Future work must bridge these gaps by developing certifiable, lightweight, and context-aware scheduling frameworks that combine the responsiveness of modern AI techniques with the rigor of classical real-time theory. As embedded systems grow increasingly intelligent and interconnected, adaptive scheduling will continue to evolve, not just as a performance enhancer, but as a foundational enabler of safe, efficient, and resilient real-time computing.
\bibliography{biblio}

@inproceedings{datta2022adaptive,
  title={Adaptive Real-Time Scheduler for Embedded Operating System},
  author={Datta, Arkajit and Rao, Shamith D. and Mohan, C. G.},
  booktitle={2nd Indian International Conference on Industrial Engineering and Operations Management},
  year={2022}
}

@article{lu2002feedback,
  title={Feedback Control Real-Time Scheduling: Framework, Modeling, and Algorithms},
  author={Lu, Chenyang and Stankovic, John A. and Son, Sang H. and Tao, Gang},
  journal={Real-Time Systems},
  volume={23},
  number={1},
  pages={85--126},
  year={2002},
  publisher={Springer}
}

@article{Gracias,
  title={Leveraging AI and ML for Predictive Scheduling in Real-Time Operating Systems},
  author={Gracias, Abram and Broklyn, Peter},
  journal={Real-Time Systems},
  year={2025}
}

@article{subramaniyan2025fc,
  title={FC-GPU: Feedback Control GPU Scheduling for Real-Time Embedded Systems},
  author={Subramaniyan, Srinivasan and Wang, Xiaorui},
  journal={ACM Transactions on Embedded Computing Systems},
  volume={24},
  number={5s},
  pages={1--25},
  year={2025},
  publisher={ACM}
}

@article{goksoy2024runtime,
  title={Runtime Monitoring of ML-Based Scheduling Algorithms Toward Robust Domain-Specific SoCs},
  author={Goksoy, A. Alper and Kanani, Alish and Chatterjee, Satrajit and Ogras, Umit},
  journal={IEEE Transactions on Computer-Aided Design of Integrated Circuits and Systems},
  volume={43},
  number={11},
  pages={4202--4213},
  year={2024},
  publisher={IEEE}
}

@article{zhao2024task,
  title={Task-Degradation Aware Adaptive Dynamic Scheduling for Priority-Based Automotive Cyber-Physical Systems},
  author={Zhao, Kaiyu and Wu, Jinming and Zou, Yuan and Zhang, Xudong and Wang, Tianyu},
  journal={IEEE Access},
  year={2024},
  publisher={IEEE}
}

@article{pan2024deep,
  title={A Deep Reinforcement Learning-Based Scheduling Framework for Real-Time Workflows in the Cloud Environment},
  author={Pan, Jiahui and Wei, Yi},
  journal={Expert Systems with Applications},
  volume={255},
  pages={124845},
  year={2024},
  publisher={Elsevier}
}

@article{he2025closed,
  title={Closed-Loop Gasoline Blending Scheduling Based on Real-Time Optimized Slack Feedback},
  author={He, Renchu and Yan, Xinyu and Hua, Junjie and Lin, Jiajiang and Zhao, Liang},
  journal={Chemical Engineering Science},
  volume={309},
  pages={121426},
  year={2025},
  publisher={Elsevier}
}

@inproceedings{posluns2025symbiotic,
  title={Symbiotic Task Scheduling and Data Prefetching},
  author={Posluns, Gilead and Jeffrey, Mark C.},
  booktitle={Proceedings of the 58th IEEE/ACM International Symposium on Microarchitecture},
  pages={140--155},
  year={2025}
}

@article{articlsnavelye,
  title={Symbiotic Job Scheduling for a Simultaneous Multithreading Machine},
  author={Snavely, Allan and Tullsen, Dean},
  journal={ACM SIGPLAN Notices},
  volume={35},
  number={9},
  pages={234--244},
  year={2000},
  doi={10.1145/356989.357011}
}

@article{abohamama2022real,
  title={Real-Time Task Scheduling Algorithm for IoT-Based Applications in the Cloud--Fog Environment},
  author={Abohamama, Abdelaziz Said and El-Ghamry, Amir and Hamouda, Eslam},
  journal={Journal of Network and Systems Management},
  volume={30},
  number={4},
  pages={54},
  year={2022},
  publisher={Springer}
}

@article{kwon2025real,
  title={Real-Time Task Scheduling and Resource Planning for IIoT-Based Flexible Manufacturing with Human--Machine Interaction},
  author={Kwon, Gahyeon and Shim, Yeongeun and Cho, Kyungwoon and Bahn, Hyokyung},
  journal={Mathematics},
  volume={13},
  number={11},
  pages={1842},
  year={2025},
  publisher={MDPI}
}

@article{kesavan2025reinforcement,
  title={Reinforcement Learning Based Secure Edge Enabled Multi Task Scheduling Model for Internet of Everything Applications},
  author={Kesavan V., Thiruppathy and R., Venkatesan and Wong, Wai Kit and Ng, Poh Kiat},
  journal={Scientific Reports},
  volume={15},
  number={1},
  pages={6254},
  year={2025},
  publisher={Nature}
}

@article{guo2025machine,
  title={When Machine Learning and Neural Networks Marry Real-Time Scheduling},
  author={Guo, Zhishan},
  journal={Real-Time Systems},
  volume={61},
  number={2},
  pages={320--325},
  year={2025},
  publisher={Springer}
}

@article{li2025real,
  title={Real-Time Data-Driven Hybrid Synchronization for Integrated Planning, Scheduling, and Execution Toward Industry 5.0 Human-Centric Manufacturing},
  author={Li, Mingxing and Ling, Shiquan and Qu, Ting and Lu, Shan and Li, Ming and Guo, Daqiang and He, Zhen and Huang, George Q.},
  journal={IEEE Transactions on Systems, Man, and Cybernetics: Systems},
  year={2025},
  publisher={IEEE}
}

@article{li2025energy,
  title={Energy-Efficient Computation with DVFS Using Deep Reinforcement Learning for Multi-Task Systems in Edge Computing},
  author={Li, Xinyi and Zhou, Ti and Wang, Haoyu and Lin, Man},
  journal={IEEE Transactions on Sustainable Computing},
  year={2025},
  publisher={IEEE}
}

@article{li2024fidrl,
  title={Fidrl: Flexible Invocation-Based Deep Reinforcement Learning for DVFS Scheduling in Embedded Systems},
  author={Li, Jingjin and Jiang, Weixiong and He, Yuting and Yang, Qingyu and Gao, Anqi and Ha, Yajun and Ozcan, Ender and Bai, Ruibin and Cui, Tianxiang and Yu, Heng},
  journal={IEEE Transactions on Computers},
  year={2024},
  publisher={IEEE}
}

@article{benmachiche2025adaptive,
  title={Adaptive Hybrid PSO--APF Algorithm for Advanced Path Planning in Next-Generation Autonomous Robots},
  author={Benmachiche, Abdelmadjid and Derdour, Makhlouf and Kahil, Moustafa Sadek and Ghanem, Mohamed Chahine and Deriche, Mohamed},
  journal={Sensors},
  volume={25},
  number={18},
  pages={5742},
  year={2025},
  publisher={MDPI}
}

@article{makhlouf2024enhanced,
  title={Enhanced Autonomous Mobile Robot Navigation Using a Hybrid BFO/PSO Algorithm for Dynamic Obstacle Avoidance},
  author={Makhlouf, Amina and Benmachiche, Abdelmadjid and Boutabia, Ines},
  journal={Informatica},
  volume={48},
  number={17},
  year={2024}
}

@inproceedings{sedraoui2024intrusion,
  title={Intrusion Detection with Deep Learning: A Literature Review},
  author={Sedraoui, Brahim Khalil and Benmachiche, Abdelmadjid and Makhlouf, Amina and Chemam, Chaouki},
  booktitle={2024 6th International Conference on Pattern Analysis and Intelligent Systems (PAIS)},
  pages={1--8},
  year={2024},
  organization={IEEE}
}

@inproceedings{benmachiche2022development,
  title={Development of a Biometric Authentication Platform Using Voice Recognition},
  author={Benmachiche, Abdelmadjid and Hadjar, Bouzata and Boutabia, Ines and Betouil, Ali Abdelatif and Maatallah, Majda and Makhlouf, Amina},
  booktitle={2022 4th International Conference on Pattern Analysis and Intelligent Systems (PAIS)},
  pages={1--7},
  year={2022},
  organization={IEEE}
}

@article{benmachiche2020optimization,
  title={Optimization Learning of Hidden Markov Model Using the Bacterial Foraging Optimization Algorithm for Speech Recognition},
  author={Benmachiche, Abdelmadjid and Makhlouf, Amina and Bouhadada, Tahar},
  journal={International Journal of Knowledge-Based and Intelligent Engineering Systems},
  volume={24},
  number={3},
  pages={171--181},
  year={2020},
  publisher={SAGE}
}

@article{benmachiche2019optimization,
  title={Optimization of Hidden Markov Model with Gaussian Mixture Densities for Arabic Speech Recognition},
  author={Benmachiche, Abdelmadjid and Makhlouf, Amina},
  journal={WSEAS Transactions on Signal Processing},
  volume={15},
  pages={85--94},
  year={2019},
  publisher={WSEAS}
}

@article{ben,
  title={Real-Time Machine Learning for Embedded Anomaly Detection},
  author={Benmachiche, Abdelmadjid and Rais, Khadija and Hamda, Slimi},
  journal={arXiv preprint arXiv:2512.19383},
  year={2025},
  doi={10.48550/arXiv.2512.19383}
}
\bibliographystyle{unsrt}

\end{document}